# Sobre la Generación de Tráfico Autosimilar con Dependencia de Largo Alcance Empleando Mapas Caóticos Unidimensionales Afines por Tramos (Versión Extendida)

# On the Generation of Self-similar with Long-range Dependent Traffic Using Piecewise Affine Chaotic One-dimensional Maps (Extended Version)

Ginno Millán Naveas[1]

**RESUMEN**

Se presenta una extensión cualitativa y cuantitativa de los modelos caóticos utilizados para generar tráfico autosimilar con dependencia de largo alcance (LRD), a través de la formulación de un modelo que considera el empleo de mapas caóticos unidimensionales afines por tramos. Sobre la base de la desagregación de las series temporales generadas se propone una explicación válida del comportamiento de los valores del exponente de Hurst y se demuestra la factibilidad de su control a partir de los parámetros del modelo propuesto.

Palabras clave: Autosimilitud, caos, exponente de Hurst, mapas caóticos, modelado de tráfico en redes de computadoras.

*ABSTRACT*

*A qualitative and quantitative extension of the chaotic models used to generate self-similar traffic with long-range dependence (LRD) is presented by means of the formulation of a model that considers the use of piecewise affine one-dimensional maps. Based on the disaggregation of the temporal series generated, a valid explanation of the behavior of the values of Hurst exponent is proposed and the feasibility of their control from the parameters of the proposed model is shown.*

## INTRODUCCIÓN

El comportamiento caótico se sitúa como un paradigma intermedio que entrelaza dos concepciones filosóficas y científicas del universo, a saber; el conocimiento absoluto regido por el determinismo y el desconocimiento total dictaminado por la aleatoriedad. Paradójicamente, y en estricta razón a la existencia de ambos dogmas, es que una aseveración que puede resultar tan natural como la anteriormente citada, no deja sino entrever la principal falencia en el análisis de comportamientos sistémicos: el uso ampliado de dicotomías para caracterizarlos.

En este escenario, es que la teoría del caos definida por Kellert [1] como el estudio cualitativo de la conducta periódica e inestable perceptible en sistemas dinámicos deterministas y no lineales, irrumpe estableciendo la impredecibilidad como característica fundamental de la experiencia común [2]. Por lo tanto, la teoría del caos en vez de tratar de entender la conducta sistémica desde un punto de vista meramente cuantitativo con el objetivo de determinar exactamente sus estados futuros, se ocupa de entender sus conductas al largo plazo explorando en la búsqueda de patrones bajo una filosofía holística, en vez de sesgarse al reductivismo que las anteriormente dichas concepciones necesariamente implican.

Como se infiere del conjunto de ideas antes expuestas, y en completa atención al ánimo de esta investigación, no resulta posible (práctico por lo demás) un abordaje de la problemática de la caracterización del comportamiento de los sistemas de interés considerando toda la amplitud conceptual de la teoría del caos. Luego, por este motivo, se acepta que el caos es el fenómeno intermedio del cual sistemas no lineales de bajo orden exhiben complejidad y un comportamiento aparentemente aleatorio [3]. Estos sistemas son de bajo orden porque pueden ser descritos correctamente por un número reducido de parámetros y variables. Se trata además de sistemas dinámicos, lo que implica que el tiempo hace evolucionar a las variables, y deterministas; esto último por cuanto los valores de las variables pueden ser, para cualquier instante de tiempo, determinados a partir de sus valores anteriores dado un conjunto de leyes dinámicas. Por último, estas leyes que rigen la evolución temporal del sistema son no lineales, es decir, incumplen el principio de superposición [4].

En este punto conviene aclarar que los sistemas caóticos difieren de los sistemas dinámicos convencionales desde la perspectiva de que son intrínsecamente impredecibles aun cuando sus leyes dinámicas subyacentes posean un carácter determinista. Además, lo anterior no debe llevar a la creencia generalizada de que el caos necesariamente implica impredecibilidad, puesto que ello no es correcto en atención a las dos principales fuentes de este último fenómeno, a saber: la inexactitud de los datos iniciales y su origen como característica inherente a cierta gama de relaciones no lineales entre variables numéricas [5]. Por lo tanto, el caos como propiedad del comportamiento de un sistema se refiere a su sensibilidad a las condiciones iniciales; es decir, dadas dos trayectorias arbitrariamente cercanas en el espacio de fases de un sistema caótico, su comportamiento será exponencialmente divergente a un ritmo dado por el exponente global de Lyapunov.

---

[1] Departamento de Ingeniería Eléctrica. Universidad de Santiago de Chile. Santiago, Chile. E-mail: ginno.millan@usach.cl

De lo antes expuesto, resulta ciertamente paradójico que un sistema en esencia determinista; con leyes dinámicas deterministas, exhiba un comportamiento caótico, puesto que la premisa básica de todos los sistemas dinámicos es que el conocimiento de las condiciones iniciales permite la determinación del comportamiento futuro del sistema. Cabe señalar que en la práctica, las condiciones iniciales solo pueden ser especificadas con precisión finita. Estas incertidumbres introducidas en las condiciones iniciales se amplían exponencialmente en el caso de los sistemas caóticos, lo cual resulta y explica lo impredecible de sus comportamientos. Luego, lo anterior implica que el caos supone la posibilidad de realizar buenas predicciones al corto plazo, pero imposibilita predicción alguna de orden práctico al largo plazo [6]. Así, sistemas muy simples, de inclusive solo un grado de libertad, como los reportados en [7] y [8], son capaces de dar lugar a comportamientos sorprendentemente complejos.

Por otra parte, con frecuencia, la noción de caos aparece ligada a la noción de fractal introducida por Mandelbrot [9], y aun cuando no existe a la fecha una demostración rigurosa de ello, las propiedades de las fractales parecen inherentes a los sistemas caóticos, por lo que tan solo en apariencia el caos y los conjuntos fractales son conceptos independientes y no vinculados [4], [10]. Sin embargo, no sin antes recordar que la dimensión fractal generaliza el concepto de dimensión a través de la introducción de valores no enteros para su especificación, hecho en [11] reportado con amplitud; sorprendentemente, todos los sistemas caóticos tienden a evolucionar asintóticamente en su espacio de fases hacia una región acotada llamada atractor extraño, la cual presenta dimensión no entera, es decir, fractal. Puede así argumentarse el hecho de que muy a menudo los atractores extraños son fractales en su naturaleza y son capaces de exhibir complejidad sobre diferentes escalas de tiempo o espacio. Luego, por todo lo anterior, es posible afirmar que los conceptos de la geometría fractal pueden ser utilizados para describir las características evolutivas de los sistemas caóticos y, a su vez, los sistemas caóticos pueden ser convenientemente utilizados para generar estructuras fractales, implicando, por lo tanto, la autosimilitud y, con ello, su parámetro de caracterización, es decir, el exponente de Hurst ($H$).

Cabe señalar que al no existir una definición sencilla para las fractales, estas, generalmente, se definen en términos de sus atributos, por ejemplo, el lento decaimiento de su varianza, los momentos de orden infinito, el ruido $1/f$, la dependencia de largo alcance (LRD), la autosimilitud, la distribución de cola hiperbólica que muestra la densidad de tiempo entre arribos sucesivos y la antes mencionada dimensión no entera, entre otros [12]-[21]. Es por tanto la presencia de estas características en los flujos de tráfico de las actuales redes de computadoras de alta velocidad el fin último de toda la discusión planteada.

El comportamiento autosimilar de los flujos de tráfico en las actuales redes de computadoras de alta velocidad, es un hecho ampliamente reportado para diferentes niveles de cobertura [22]-[35], tecnologías de transmisión [36]-[46], protocolos de control y de señalización [47], [48] y aplicaciones, en especial de audio y video [49]-[54]. Así mismo, el problema del modelado del tráfico de entrada a las redes de comunicaciones es un tema extensamente tratado en la literatura que ha dado origen a una amplia variedad de propuestas, entre las cuales se encuentran, sin ánimo de construir una lista exhaustiva, los Procesos de Poisson con Conmutación Generalizada (GSPP) [55], los Procesos de Markov Modulados por Poisson (MMPP) [56], los Procesos de Poisson Conmutados (SPP) [57], los Procesos Fractales Puntuales (FPP) [58], [59], los Procesos con Renovación Fractal Alternante (AFRP) y su variante Alternante Extendida (EAFRP) [23], [60], los procesos basados en la iteración de mapas caóticos intermitentes [3], [4], [61]-[63] y los tradicionales de ruido Gaussiano fraccional (fGn) y de movimiento Browniano fraccional (fBm) [64]-[67].

Sin embargo, a pesar de todos los esfuerzos subyacentes a cada una de las metodologías expuestas, se presentan dos situaciones problemáticas inherentes a la generación de tráfico autosimilar LRD: el grado de representatividad que posee el exponente de Hurst como parámetro único para caracterizar el rendimiento de los sistemas de colas en los que se presenta y el comportamiento que su valor exhibe cuando se lleva a cabo una desagregación de las series temporales de segundo orden autosimilares que se obtienen dentro del intervalo de interés $0.5 < H < 1$. En [68]-[71] se presentan aisladamente ambos problemas y sus implicancias. En esta investigación, al conjunto de ambas situaciones y en particular al de sus repercusiones sobre los sistemas bajo estudio se le denomina *localidad del exponente de Hurst (H)*.

En atención al conjunto existente de modelos de tráfico que utilizan mapas caóticos, se presenta una extensión de los mismos incorporando a los mapas afines por tramos para desarrollar un modelo de tráfico fractal que además proporciona una explicación a la localidad de $H$ para el tráfico generado y mitiga sus efectos.

## MAPAS CAÓTICOS Y TRÁFICO AUTOSIMILAR

El uso de mapas caóticos para modelar fuentes de tráfico fue propuesto por primera vez en [3], [4], [48], sobre las bases propuestas por el trabajo pionero [72]. En esencia, un mapa caótico puede comprenderse como una variante del modelo de tráfico On/Off descrito y desarrollado en [23], con la diferencia fundamental de que su base ya no radica en los tradicionales enfoques probabilísticos, sino más bien en las dinámicas discretas que subyacen tras el comportamiento del sistema.

Sea un sistema caótico caracterizado por una función no lineal $f: D \to D$, con $D \subseteq \mathbb{R}^m$. El sistema evoluciona de acuerdo con la ecuación del proceso, definida por $f$, que para un sistema discreto viene dada por

$$\mathbf{x}[k] = f(\mathbf{x}[k-1], u[k-1]; \boldsymbol{\theta}) + \mathbf{v}[k], \qquad (1)$$

con $\mathbf{x}[k]$ vector de estado del sistema, $u[k]$ la excitación de entrada al sistema, $\mathbf{v}[k]$ vector de ruido del proceso y $\boldsymbol{\theta}$ vector de parámetros del sistema.

En general, el estado del sistema no puede ser observado en forma directa, hecho por el cual se requieren muestras del mismo, las cuales se obtienen mediante la aplicación de un cierto proceso de medida que puede ser definido a través de

$$\mathbf{y}[k] = g(\mathbf{x}[k]) + \mathbf{w}[k], \quad \text{con } k = 0, \ldots, N, \qquad (2)$$

donde $g$ es la función de medida (la cual puede ser o no conocida), $\mathbf{y}[k]$ el vector de medidas y $\mathbf{w}[k]$ el vector de ruido de medida.

Un mapa caótico es una aplicación $f: X \to Y$ que asocia cada $x \in X \subseteq \mathbb{R}^m$ con un único $y \in Y \subseteq \mathbb{R}^q$, siendo $X$ el dominio de $f$ e $Y$ el conjunto de llegada de $f$.

Sea una aplicación $f: D \to D$, con $D \subseteq \mathbb{R}^m$. Se dice que un mapa iterado es el sistema formado por el conjunto de $m$ ecuaciones dado por

$$\mathbf{x}[n] = f(\mathbf{x}[n-1]; \boldsymbol{\theta}). \qquad (3)$$

Las ecuaciones (1)-(3) muestran que un mapa caótico es simplemente una clase de sistema discreto autónomo.

Puesto que este trabajo centra su atención en los mapas caóticos unidimensionales, es conveniente establecer su definición como sigue: un mapa caótico unidimensional es una aplicación $f: D \to D$, con $D \subseteq \mathbb{R}$, tal que

$$x[n] = f(x[n-1]; \theta). \qquad (4)$$

La definición del caos como propiedad de un sistema se refiere a su sensibilidad a las condiciones iniciales. De esta manera, considerando un mapa caótico definido por $x_{n+1} = f(x)$ y dos trayectorias arbitrarias con condiciones iniciales casi idénticas dadas por $x_0$ y $x_0 + \varepsilon$, con $\varepsilon \to 0$, su sensibilidad a las condiciones iniciales se define por

$$\left| f^N(x_0 + \varepsilon) - f^N(x_0) \right| = \varepsilon \exp[N \lambda(x_0)] \qquad (5)$$

donde $f^N(\cdot)$ representa a la $N$-ésima iteración del mapa y $\lambda(x_0)$ es el exponente global de Lyapunov que describe la divergencia exponencial.

Para que el mapa $x_{n+1} = f(x)$ sea caótico, el exponente global de Lyapunov debe ser positivo para la mayoría de los $x_0$ [73]. Además, (5) reafirma esta condición, puesto que permite argumentar que puntos que comienzan con condiciones iniciales similares, se desarrollan a lo largo de trayectorias diferentes.

Sea una aplicación $f: I \to I$, con $I \in \mathbb{R}$. Se dice que $f(x)$ es un mapa unidimensional afín por tramos, si existe un número finito de puntos $e_0 < e_1 < \ldots e_M$, de modo que el intervalo $I = [e_0, e_M]$ puede subdividirse en $M$ intervalos menores dados por $E_i = [e_{i-1}, e_i)$, con $i = 1, \ldots M - 1$ y $E_M = [e_{M-1}, e_M]$, dentro de los cuales el comportamiento de $f(x)$ es afín.

Matemáticamente, un mapa caótico unidimensional afín por tramos se expresa como

$$f(x) = \sum_{i=1}^{M} (a_i x + b) \Psi_{E_i}(x), \qquad (6)$$

donde $\Psi$ es la función característica o indicadora, que se define como sigue. Sea una aplicación $\Psi_R: D \to \{0, 1\}$ con $D \in \mathbb{R}$. Se dice que $\Psi_R$ es la función característica o indicadora de la región $R$, si presenta la forma

$$\Psi_R = \begin{cases} 1, & x \in R \\ 0, & x \notin R \end{cases}. \qquad (7)$$

En otras palabras, dado el mapa caótico unidimensional afín por tramos $f(x)$, la función característica $\Psi$ obliga a $f(x)$ a presentar un comportamiento semejante dentro de cada uno de los $M$ intervalos $E_i$ en los cuales se divide el intervalo $I$ de llegada, que si se observase a $f(x)$ dentro del intervalo $I$ en su totalidad. Este hecho evidencia que el mapa $f(x)$ posee un comportamiento autosimilar.

Un mapa unidimensional, compuesto por dos intervalos, en el cual la variable de estado $x_n$ evoluciona en el tiempo de acuerdo con dos funciones $f_1(\cdot)$ y $f_2(\cdot)$ que satisfacen la condición (5), se puede escribir a partir de (4), como

$$x_{n+1} = \begin{cases} f_1(x_n), & 0 < x_n \le d \\ f_2(x_n), & d < x_n < 1 \end{cases}. \quad (8)$$

Esta notación posibilita la concepción de un proceso de generación de tramas considerando una fuente de origen que alterna entre en un estado pasivo y un estado activo (de manera similar al caso de los modelos On/Off) en un instante determinado $n$, en función de si el valor de la variable de estado $x_n$ se encuentra sobre o debajo de un cierto umbral de activación $d$. De esta manera, todas las iteraciones del mapa en el estado activo corresponden a procesos de generación de tramas y todas las iteraciones del mapa en el estado pasivo corresponden a procesos de tiempo entre arribos sucesivos.

Bajo la misma lógica de razonamiento y considerando la permanencia del mapa en uno u otro de los dos estados anteriores producto de la activación de una u otra de sus funciones componentes, la evolución de los procesos de llegadas sucesivas de tramas queda definida a partir de la función indicadora (7), la cual por consistencia con la notación de (8), puede ser escrita como

$$y_n = y(n) = \begin{cases} 0, & 0 < x_n \le d \\ 1, & d < x_n < 1 \end{cases}. \quad (9)$$

La referencia [74] reporta una interesante interpretación del modelo descrito por (8) y (9), considerándolo como la unión de dos capas dinámicas; una oculta dada por $x_n$ y una visible especificada por $y_n$.

Es importante notar que el comportamiento de cualquier trayectoria sobre la cual evoluciona el mapa descrito por (8), es tal que no tiene porqué visitar las dos regiones de su espacio de fases con la misma frecuencia; es más, no existe razón alguna para considerar, inclusive dentro del propio atractor, una función de densidad de probabilidad uniforme de las secuencias generadas. Resulta razonable entonces preguntarse tanto por la frecuencia con la cual una determinada trayectoria visita cada región del mapa en un periodo de observación de $n$ iteraciones, como por la forma de lograr calcular dicha función de densidad de probabilidad a partir de una condición inicial $x_0$ dada. Al respecto, la respuesta a ambas interrogantes se halla en l distribución de densidad de estados del mapa, $\rho_n(x)$, que de acuerdo a [75] está dada por

$$\rho_n(x) = \frac{1}{N} \sum_{i=1}^{N} \delta[x - x_n(i)], \quad (10)$$

donde $\delta(x)$ es la función delta de Dirac y la evolución de $\rho_n(x)$ es definida por la ecuación de Frobenius-Perron, la cual, de [76], viene dada por la expresión

$$\rho_{n+1}(x) = \int \delta[x - f(z)] \rho_n(z) dz. \quad (11)$$

Para el caso de un mapa unidimensional $x_{n+1} = f(n)$ con $x_n \in [0, 1]$ tal que $n \in \mathbb{N}_0$, de (10) se tiene que

$$\rho(x) = \lim_{N \to \infty} \frac{1}{N} \sum_{n=0}^{N} \delta[x - f_n(x_0)]. \quad (12)$$

Si $\rho(x)$ no depende de $x_0$, el sistema es ergódico. Luego, con esta condición, en [75] se demuestra que

$$\lim_{N \to \infty} \frac{1}{N} \sum_{i=0}^{N} g(x_i) \equiv \lim_{N \to \infty} \frac{1}{N} \sum_{i=0}^{N} g[f^i(x_0)] \\ = \int_0^1 \rho(x) g(x) dx. \quad (13)$$

Sin embargo, dado que $\rho_n(x)$ tiene que ser debido a (13), estacionaria, en otras palabras no depende del instante $n$, se conoce como densidad invariante del mapa $f(x)$ [48] y describe la densidad de iteraciones de $x_n$ en el intervalo $(0, 1)$ cuando $n \to \infty$. Luego, $\rho_n(x)$ es una autofunción del operador de Frobenius-Perron con autovalor 1 y, por lo tanto, se cumple que

$$\rho(x) = \int_0^1 \delta[x - f(z)] \rho(z) dz. \quad (14)$$

Una exposición precisa del conjunto de hechos que dan origen al tratamiento del tráfico en las actuales redes de computadoras de alta velocidad tomando en cuenta su naturaleza autosimilar con LRD, conlleva a la necesidad de capturar sus fluctuaciones sobre diferentes escalas de tiempo con el fin de realizar pronósticos asertivos sobre su rendimiento. Al respecto, la autosimilaridad con LRD aporta la parsimonia necesaria para el tratamiento de los detalles estadísticos de las variables involucradas a partir de un conjunto mínimo de parámetros de modelado, y un modelo desarrollado sobre la teoría de la complejidad, la robustez necesaria para cohesionar dichos parámetros.

## ESPECIFICACIÓN DEL MODELO CAÓTICO

Por simplicidad, la deducción del modelo propuesto se aborda como a continuación prosigue.

Considérese, en un primer término, el mapa no lineal de doble intermitencia originalmente propuesto en [4]

$$x_{n+1} = \begin{cases} \varepsilon_1 + x_n + c_1 x_n^{m_1}, & 0 < x_n \le d \\ -\varepsilon_2 + x_n - c_2(1-x_n)^{m_2}, & d < x_n < 1 \end{cases}, \quad (15)$$

donde

$$c_1 = (1 - \varepsilon_1 - d) / d^{m_1} \quad (16)$$

y

$$c_2 = (d - \varepsilon_2) / (1 - d)^{m_2}, \quad (17)$$

con una función indicadora dada por (9).

Si $m_2 = 1$ y $\varepsilon_2 = 0$, de (15) se obtiene el modelo de un mapa caótico intermitente no lineal [3], es decir,

$$x_{n+1} = \begin{cases} \varepsilon + x_n + c x_n^m, & 0 < x_n \leq d \\ (x_n - d)/(1 - d), & d < x_n < 1 \end{cases}, \quad (18)$$

donde

$$c = \frac{1 - \varepsilon - d}{d^m}, \quad \text{con } \varepsilon \ll d. \quad (19)$$

Los parámetros $\varepsilon$, $m$ y $d$ del mapa definido por (18) se utilizan para controlar la probabilidad de permanencia en el estado inactivo, el grado de autosimilitud ($H$) y la carga de tráfico; es decir, la tasa media de arribo de tramas. En lo específico, el ajuste de $\varepsilon$ incide sobre la probabilidad de permanencia de las iteraciones del mapa en la región de inactividad, mientras que la carga de tráfico depende de los parámetros $m$ y $d$ [77], [78].

La condición $\varepsilon \ll d$ queda bien definida, a lo menos de forma teórica, si $\varepsilon = 0$, lo cual redunda en el control del límite del rango de escalas temporales sobre las cuales se observa la LRD. En [77] se demuestra que si $\varepsilon = 0$, el tiempo de permanencia puede tener cualquier longitud, pero a medida que $\varepsilon$ se incrementa sobre 0, el tiempo de escape de la región tiende a un límite fijo superior.

Considerar $m = 1$ con el objetivo de disminuir el grado de la función que caracteriza al estado inactivo del mapa implica generar tráfico con dependencia de corto alcance (SRD). Este último hecho puede verificarse a partir de la relación existente entre $H$ y $m$; $H = (3m - 4) / (2m - 2)$, con $m = \max \{m_1, m_2\}$ [79]. En otras palabras, se produce un decaimiento geométrico para las regiones del mapa, lo cual equivale a tráfico no correlacionado, es decir tráfico con dependencia de corto alcance.

Luego, el modelo de mapa caótico propuesto junto con su función indicadora vienen dados por

$$x_{n+1} = \begin{cases} x_n + x_n^m, & 0 < x_n \leq d \\ (x_n - d)/d^m, & d < x_n < 1 \end{cases} \quad (20)$$

y

$$y_n = y(n) = \begin{cases} 0, & 0 < x_n \leq d \\ 1, & d < x_n < 1 \end{cases}, \quad (21)$$

respectivamente.

Reescribiendo (20) en la forma

$$x_{n+1} = \begin{cases} f_1(x) = x_n + x_n^m, & 0 < x_n \leq d \\ f_2(x) = (x_n - d)/d^m, & d < x_n < 1 \end{cases}, \quad (22)$$

del desarrollo de (14) aplicando (22) se tiene que

$$\rho(x) = \int_0^d \delta[x - f_1(z)] \rho(z) dz + \int_d^1 \delta[x - f_2(z)] \rho(z) dz, \quad (23)$$

ecuación para la cual en [73] se establece la existencia de sólo una solución relevante físicamente que se obtiene a partir de $z = 1 / f_i(x)$, es decir

$$\rho(x) = \frac{\rho[f_1^{-1}(x)]}{f_1'[f_1^{-1}(x)]} + \frac{\rho[f_2^{-1}(x)]}{f_2'[f_2^{-1}(x)]}, \quad (24)$$

donde $f_1'$ es la primera derivada de $f_1$.

Empleando la notación de [80] en (24), se obtiene que la expresión para la densidad invariante está dada por la expresión

$$\rho(x) = \sum_{i=1}^{2} \frac{\rho(y_i)}{f_i'(y_i)}, \quad \text{con } y_i = f_i'(x), \quad (25)$$

la cual permite obtener la carga del sistema, es decir la probabilidad de permanencia en el estado activo, a partir de la integración de (25) entre los límites $d$ y 1.

## RESULTADOS EXPERIMENTALES

El comportamiento autosimilar con LRD que muestra el tráfico generado por el modelo propuesto se comprueba mediante el cálculo de exponente de Hurst ($H$) a partir de los análisis de rango reescalado (R/S) [63], [81] y de varianza agregada (Var) [82], [83]. En el análisis R/S el valor de $H$ se obtiene directamente de la pendiente de la recta resultante del gráfico log-log de $H$ versus el nivel de agregación, mientras que en el análisis Var se obtiene de la relación $H = 1 - \beta / 2$, con $0 < \beta < 1$ [1], [84], [85].

En esta primera etapa de la investigación, la atención se centra en validar el modelo propuesto sobre la base de simulaciones. Por este motivo, se programan (20) y (21) en lenguaje MATLAB y se llevan a cabo experimentos numéricos tendientes al cálculo del parámetro de Hurst tanto de series de tráfico generadas directamente por el modelo, como de series de tráfico originadas a partir de la desagregación aleatoria de las primeras.

La Tabla 1 muestra valores del parámetro de Hurst (*H*) considerando los análisis de rango reescalado (R/S) y de varianza agregada (Var) para diferentes experimentos desarrollados para una cantidad fija de 1000 iteraciones (*N*), variando los parámetros *m* y *d* (la frontera entre las regiones de iteración del mapa).

En la Tabla 2 se muestran valores de *H* para secuencias de tráfico creadas a partir de la extracción aleatoria de 500 muestras de las series utilizadas para formular la Tabla 1. Las series resultantes se analizan como tráfico generado directamente por el modelo propuesto.

Las Tablas 3 y 4 relacionan experimentos numéricos de series originales y desagregadas, respectivamente, con sus representaciones gráficas, las cuales se muestran en el Apéndice A.

## DISCUSIÓN DE RESULTADOS

El modelo de mapa caótico propuesto demuestra que la generación de tráfico autosimilar con LRD es factible de ser abordada a partir del control de los parámetros m y d considerando una cantidad fija de iteraciones; lo cual queda en evidencia a partir de los resultados expuestos en la Tabla 1. Sin embargo, producto de lo simplificada de su actual formulación, no es posible el control de los tiempos de permanencia en cada región y, por ello, en cada estado On u Off. Al respecto, un modelo general como el reportado en [48] permite tal control con buena exactitud. Un ejemplo simple que ilustra esta necesidad se construye sobre la base de requerir que el tráfico de un sistema presente un determinado *H*, por ejemplo 0.9. al respecto, verificando las columnas *d* = 0.5 y *d* = 0.3 de la Tabla 1, se comprueba que este requerimiento solo puede ser satisfecho a partir de una cambio del límite de las regiones de iteración, hecho por demás poco práctico por sus repercusiones sobre la función indicadora dada por (7). Luego, se trata, por tanto, de una necesidad que requiere de atención urgente.

Por otra parte, el control del nivel de autosimilitud del tráfico generado por el modelo se precia de ser efectivo para las series temporales subyacentes consideradas.

Tabla 1. Experimentos con los parámetros de modelado *N*, *m* y *d* para calcular *H* de las series originales.

| *N* | *m* | *d* = 0.5 | | *d* = 0.3 | | *d* = 0.1 | |
|---|---|---|---|---|---|---|---|
| | | *H* | | *H* | | *H* | |
| | | R/S | Var | R/S | Var | R/S | Var |
| 1000 | 1 | 0.50 | 0.42 | 0.50 | 0.44 | 0.51 | 0.44 |
| | 1.2 | 0.51 | 0.46 | 0.51 | 0.47 | 0.53 | 0.50 |
| | 1.4 | 0.54 | 0.50 | 0.61 | 0.56 | 0.66 | 0.56 |
| | 1.6 | 0.68 | 0.58 | 0.72 | 0.65 | 0.75 | 0.65 |
| | 1.8 | 0.73 | 0.68 | 0.78 | 0.72 | 0.81 | 0.72 |
| | 2 | 0.87 | 0.78 | 0.90 | 0.82 | 0.92 | 0.82 |

Tabla 2. Experimentos con los parámetros de modelado *N*, *m* y *d* para calcular *H* de las series desagregadas.

| *N Aleatorio* | *m* | *d* = 0.5 | | *d* = 0.3 | | *d* = 0.1 | |
|---|---|---|---|---|---|---|---|
| | | *H* | | *H* | | *H* | |
| | | R/S | Var | R/S | Var | R/S | Var |
| 500 | 1 | 0.42 | 0.37 | 0.44 | 0.38 | 0.43 | 0.38 |
| | 1.2 | 0.45 | 0.43 | 0.47 | 0.45 | 0.45 | 0.42 |
| | 1.4 | 0.51 | 0.48 | 0.53 | 0.51 | 0.52 | 0.50 |
| | 1.6 | 0.55 | 0.52 | 0.62 | 0.57 | 0.57 | 0.52 |
| | 1.8 | 0.65 | 0.58 | 0.67 | 0.62 | 0.63 | 0.57 |
| | 2 | 0.72 | 0.66 | 0.75 | 0.67 | 0.71 | 0.66 |

Tabla 3. Parámetros para series originales.

| *N* | *d* | *m* | En Figura | Tráfico generado |
|---|---|---|---|---|
| 1000 | 0.5 | 1 | Figura 1 | |
| | | 2 | Figura 2 | Figura 9 |
| | 0.3 | 1 | Figura 3 | |
| | | 2 | Figura 4 | |

Tabla 4. Parámetros para series desagregadas.

| *N* | *d* | *m* | En Figura | Tráfico generado |
|---|---|---|---|---|
| 1000 | 0.5 | 1 | Figura 5 | |
| | | 2 | Figura 6 | Figura 10 |
| | 0.3 | 1 | Figura 7 | |
| | | 2 | Figura 8 | |

En relación con lo anterior, el análisis de los resultados de la Tabla 2 refleja la problemática de fondo: controlar el efecto de la localidad del exponente de Hurst. En este contexto, se observa que las trazas formadas a partir de tramos aleatorios de las muestras originales reflejan el comportamiento del total y, que además, el exponente de Hurst puede considerarse como un indicador válido para caracterizar los efectos del tráfico autosimilar con LRD, sobre el rendimiento de los sistemas de colas en los que se presenta. Se reconoce que no se dispone de una demostración matemática formal de este hecho; no obstante, en una primera instancia, en virtud del efecto que posee la función indicadora sobre la percepción del sistema, si se encuentra una fundamentación adecuada.

## CONCLUSIONES

Se ha presentado una extensión cualitativa y cuantitativa de los modelos de generación de tráfico autosimilar con dependencia de largo alcance incorporando los mapas caóticos afines por tramos a través de la propuesta de un nuevo modelo que proporciona una interpretación a la localidad de *H* a partir del análisis del tráfico generado.

Se comprueba la factibilidad de contar con un generador de tráfico autosimilar con dependencia de largo alcance eficiente y eficaz a partir de la parsimonia de su modelo; hecho que queda en evidencia por un control adecuado del valor del exponente de Hurst.

Se demuestra que el exponente de Hurst puede ser, en si mismo, ser considerado como un parámetro válido para caracterizar tráfico telemático autosimilar.

Se demuestra que el valor que el valor que el exponente de Hurst muestra luego de la desagregación de las series temporales autosimilares depende del modelo de origen del tráfico.

Respecto de lo anterior, sobre subconjuntos de muestras obtenidas aleatoriamente a partir de las series originales, se observa que su valor presenta, con claridad, tendencia a mantenerse y caracterizar así de forma adecuada a los segmentos en cuestión.

Finalmente se señala que el hecho más relevante para el desarrollo futuro de la investigación, es la demostración de que $m = 1$ debe ser descartado como una opción para formular un modelo caótico representativo de flujos de tráfico autosimilar, puesto que conlleva inevitablemente a tráfico con dependencia de corto alcance, lo cual queda en evidencia con los resultados expuestos en las Tablas 1 y 2. Luego, se determina la imposibilidad de trabajar con un modelo lineal y con ellos la imposibilidad de que el modelo propuesto pueda ser puramente afín por tramos como originalmente se propuso demostrar.

# ANEXO A

Las Figuras 1 a 4 exhiben el comportamiento de $x_{n+1}$ e $y_n$ para los datos de la Tabla 3. La Figura 1(b) muestra que solo hasta $N = 51$ $y_n$ alterna entre 0 y 1. Las Figuras 5 a 8 muestran $x_{n+1}$ e $y_n$, para los datos de la Tabla 4.

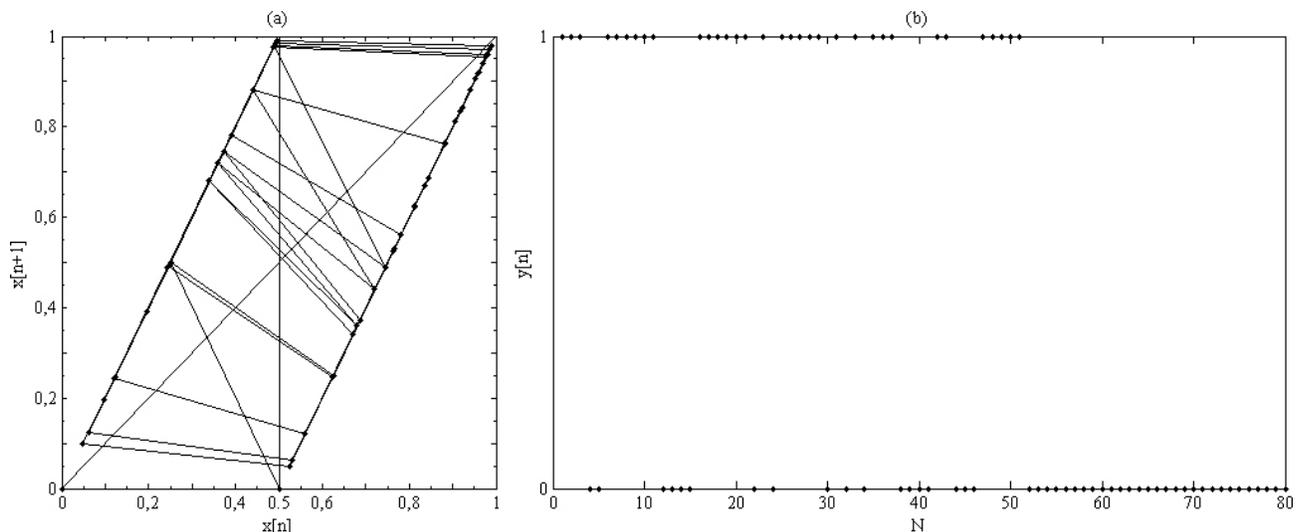

Figura 1. Comportamiento del modelo considerando $N = 1000$ iteraciones, $d = 0.5$ y $m = 1$.

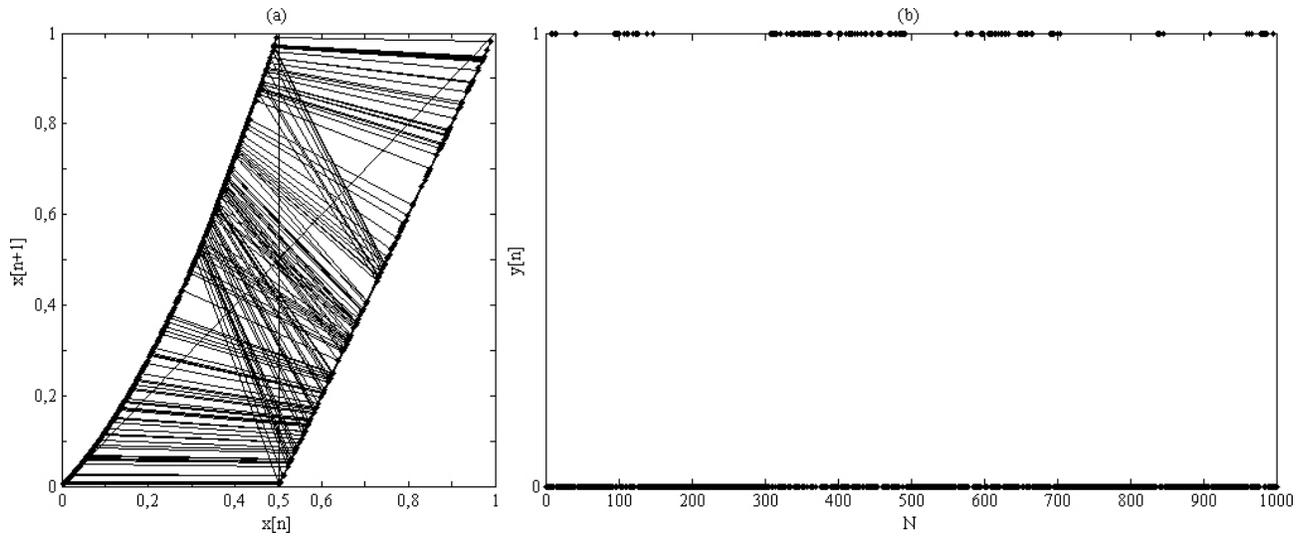
Figura 2. Comportamiento del modelo considerando $N = 1000$ iteraciones, $d = 0.5$ y $m = 2$.

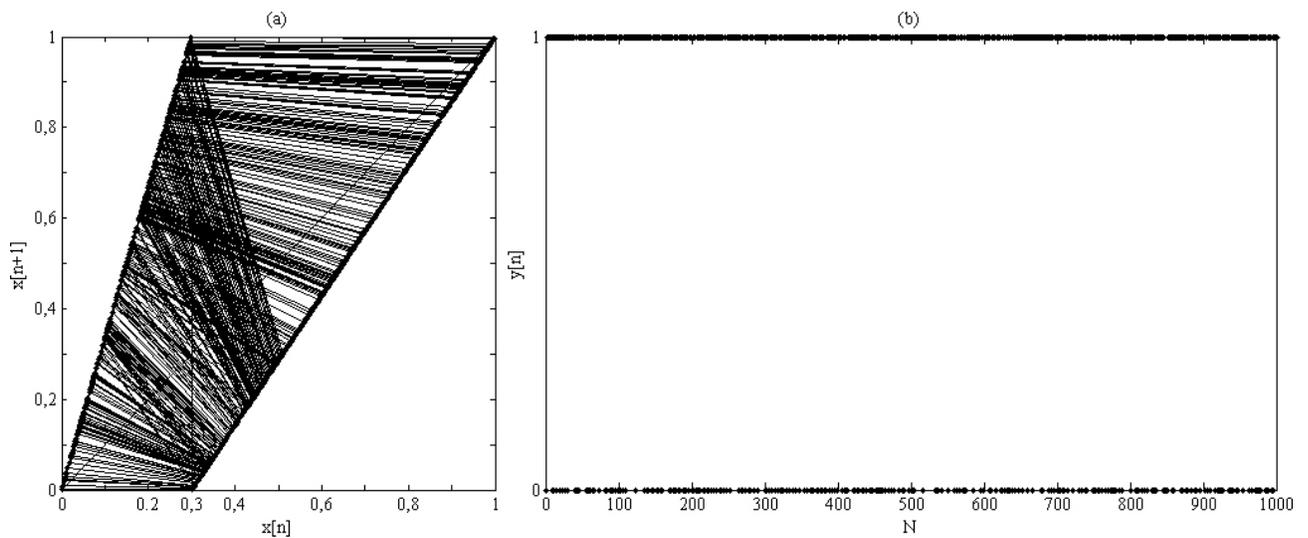
Figura 3. Comportamiento del modelo considerando $N = 1000$ iteraciones, $d = 0.3$ y $m = 1$.

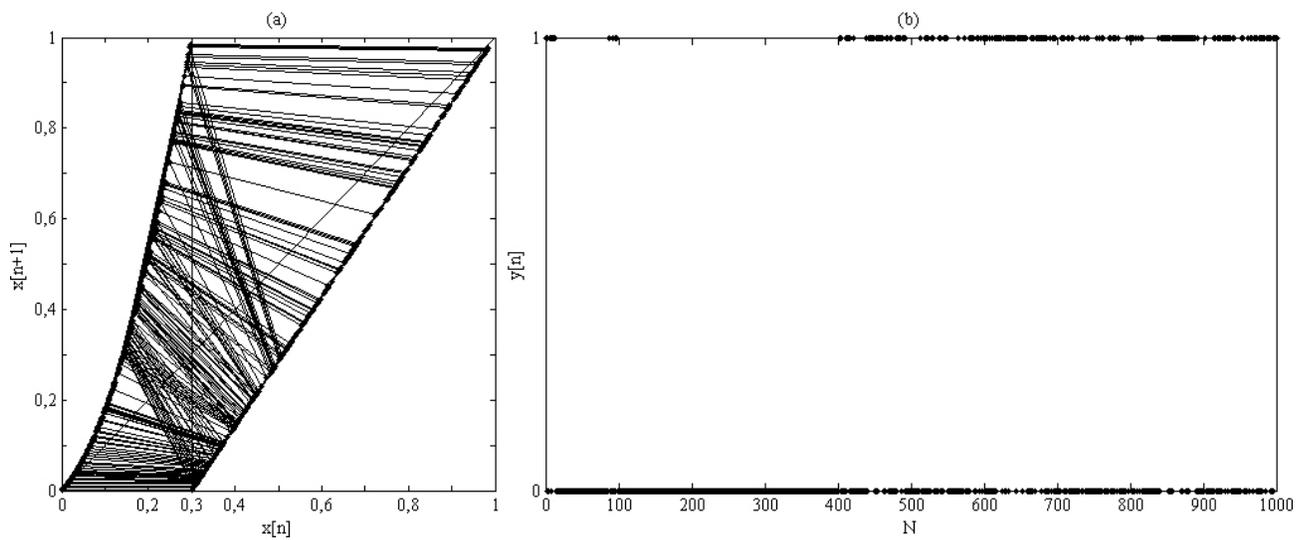
Figura 4. Comportamiento del modelo considerando $N = 1000$ iteraciones, $d = 0.3$ y $m = 2$.

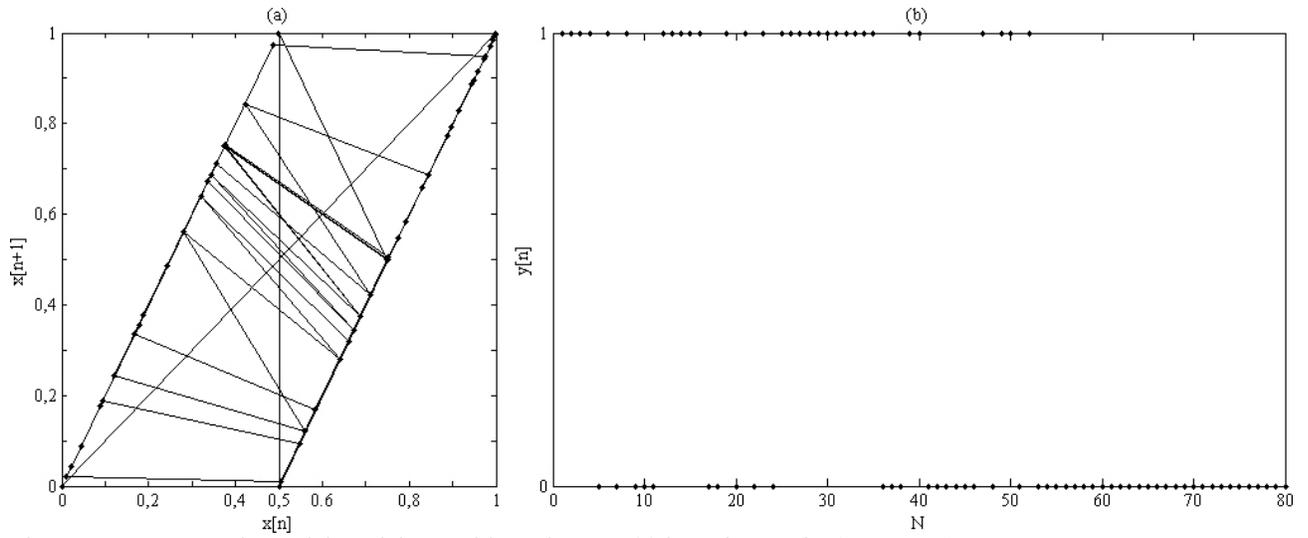
Figura 5. Comportamiento del modelo considerando $N = 500$ iteraciones, $d = 0.5$ y $m = 1$.

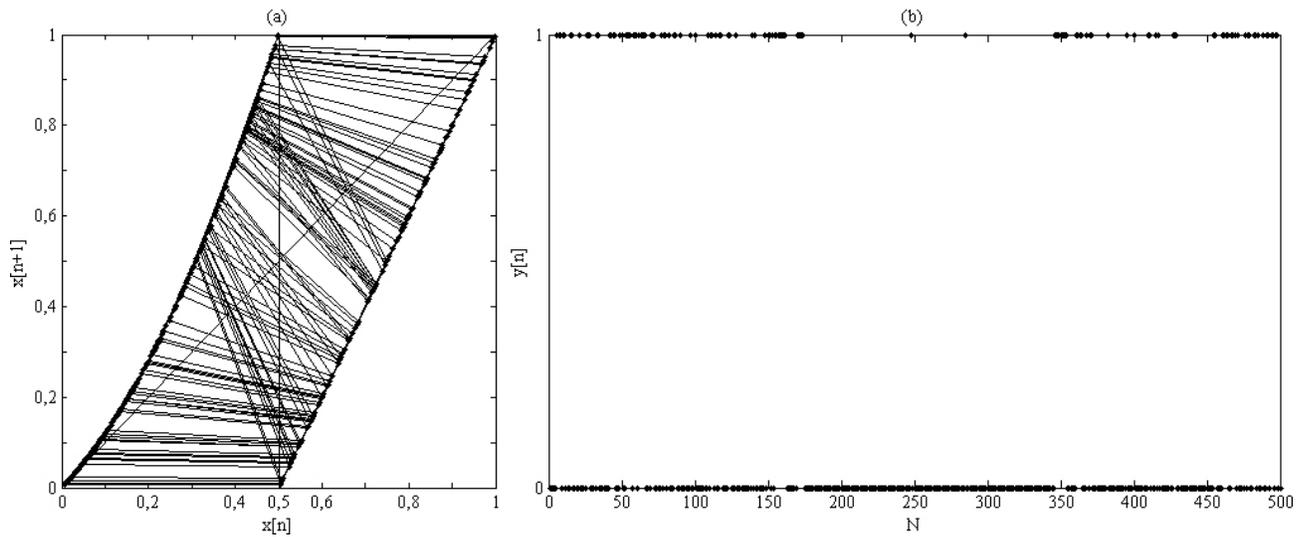
Figura 6. Comportamiento del modelo considerando $N = 500$ iteraciones, $d = 0.5$ y $m = 2$.

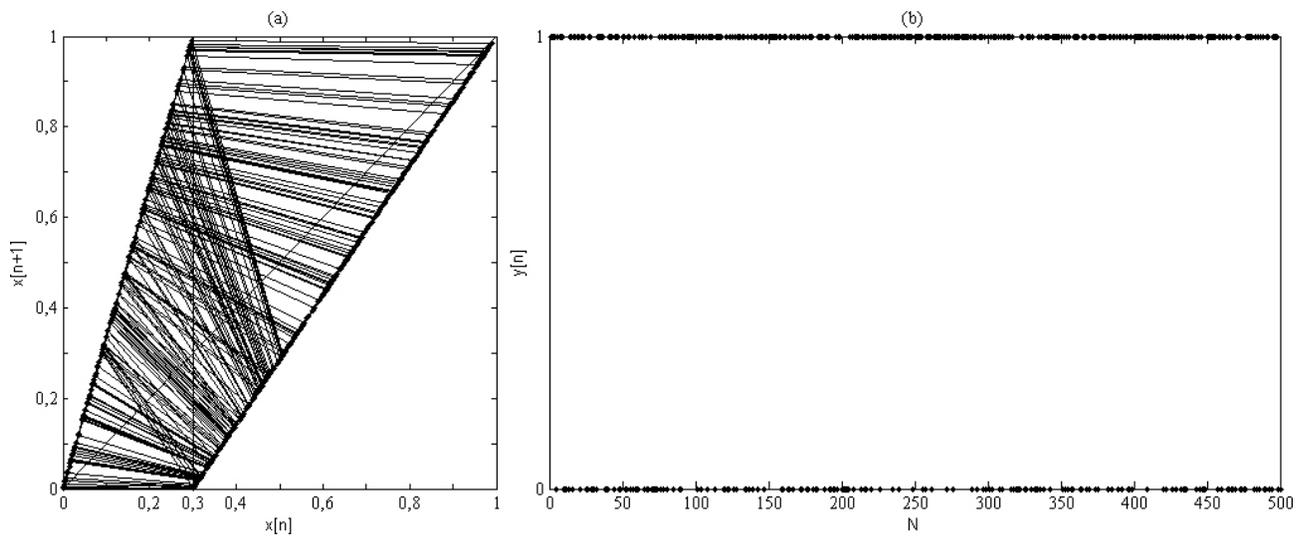
Figura 7. Comportamiento del modelo considerando $N = 500$ iteraciones, $d = 0.3$ y $m = 1$.

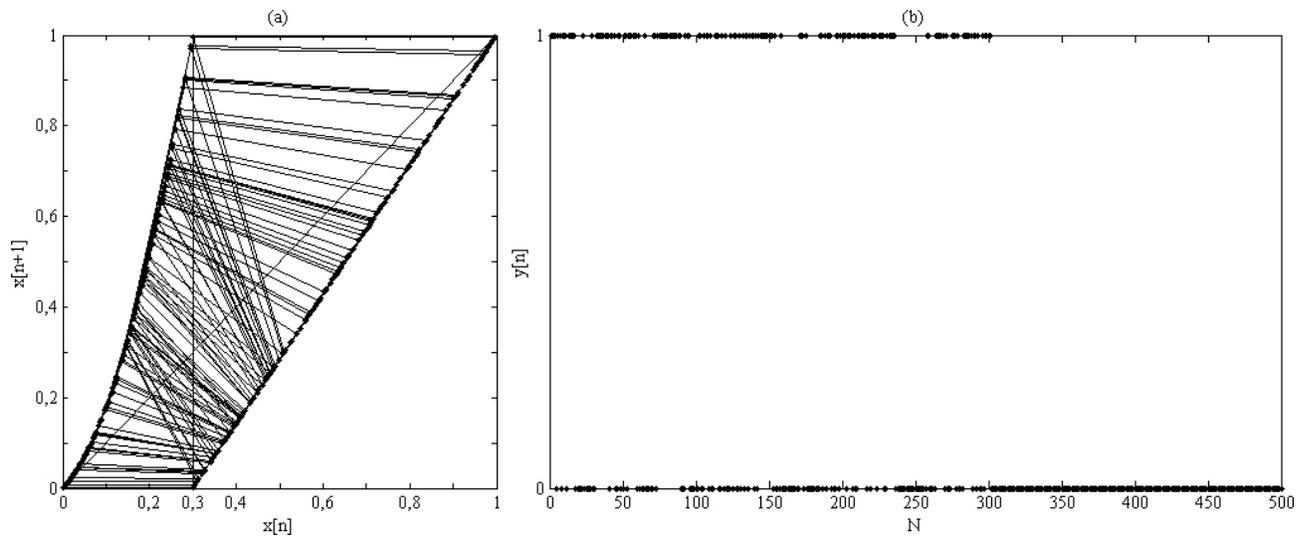
Figura 8. Comportamiento del modelo considerando $N = 500$ iteraciones, $d = 0.3$ y $m = 2$.

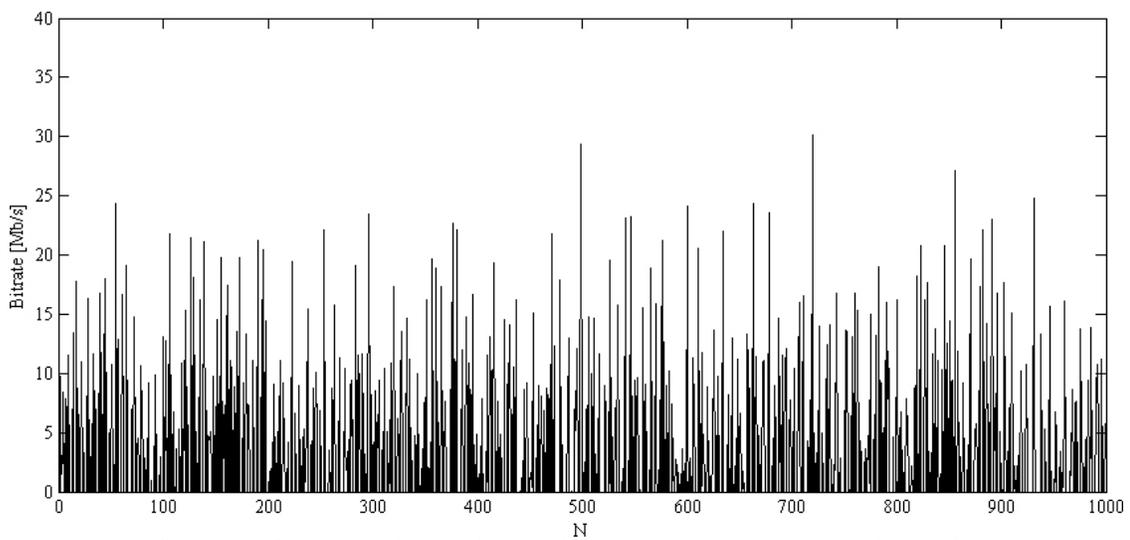
Figura 9. Aspecto gráfico del tráfico generado por el mapa de parámetros $N = 1000$ iteraciones, $d = 0.5$ y $m = 2$.

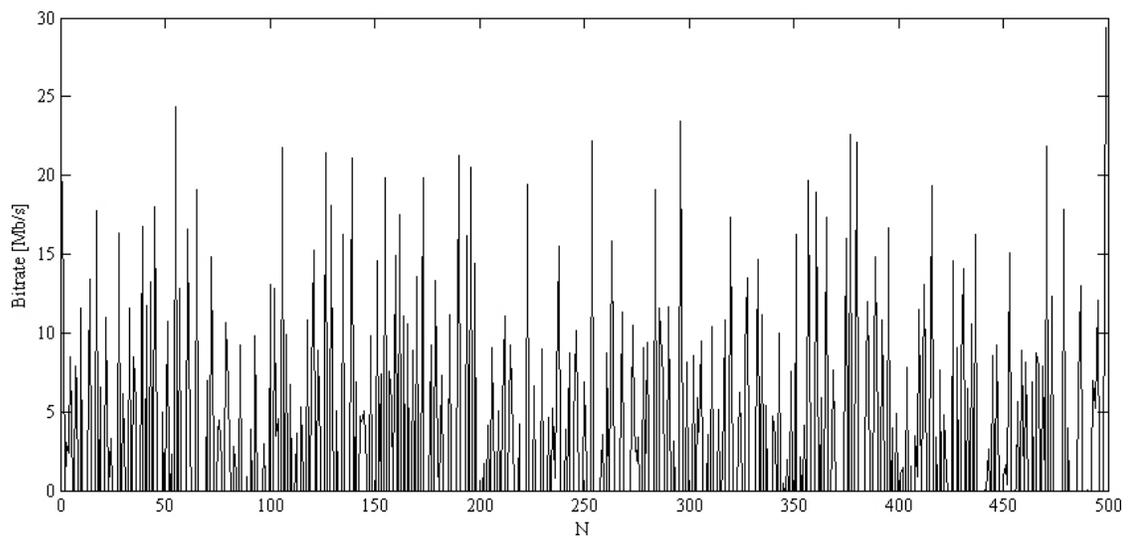
Figura 10. Aspecto gráfico del tráfico generado por el mapa de parámetros $N = 1000$ iteraciones, $d = 0.5$ y $m = 2$.